%% file: main.tex
\documentclass[compsoc, conference, a4paper, 10pt, times]{IEEEtran}

\usepackage[hidelinks]{hyperref}
\usepackage{booktabs} 
\usepackage{siunitx}
\usepackage{graphicx} 
\usepackage[ruled,vlined]{algorithm2e}
\usepackage{float}

\begin{document}

\title{Using Honeybuckets to Characterize Cloud Storage Scanning in the Wild}

\author{
\IEEEauthorblockN{Katherine Izhikevich}
\IEEEauthorblockA{University of California San Diego}\\   
\IEEEauthorblockN{Stefan Savage}
\IEEEauthorblockA{University of California San Diego}\\ 
\and
\IEEEauthorblockN{Geoff Voelker}
\IEEEauthorblockA{University of California San Diego}\\            
\IEEEauthorblockN{Liz Izhikevich}
\IEEEauthorblockA{Stanford University}
}

\maketitle

\begin{abstract}
In this work, we analyze to what extent actors target poorly-secured cloud storage buckets for attack. 
We deployed hundreds of AWS S3 honeybuckets with different names and content to lure and measure different scanning strategies. 
Actors exhibited clear preferences for scanning buckets that appeared to belong to organizations, especially commercial entities in the technology sector with a vulnerability disclosure program. 
Actors continuously engaged with the content of buckets by downloading, uploading, and deleting files.
Most alarmingly, we recorded multiple instances in which malicious actors downloaded, read, and understood a document from our honeybucket, leading them to attempt to gain unauthorized server access. 
\end{abstract}

\maketitle
\input{intro}
\input{related}
\input{phase1.tex}
\input{phase2.tex}
\input{ethics.tex}
\input{discussion}
\input{conclusion.tex}
\input{data_avail}
\bibliographystyle{plain}
\bibliography{references}
\newpage
\input{appendix.tex}
\end{document}

%% file: intro.tex
\section{Introduction}
This paper explores a simple but poorly understood question:
\emph{to what extent is insecure cloud storage actively targeted for attack?}

Storage in the cloud, such as Amazon Simple Storage Service (S3) and Google Cloud Storage, provides reliable, available, and elastic storage on demand to virtually anyone with the means to pay.  Moreover, it is extremely easy to deploy.  A client need only choose a name for their storage ``bucket'' and specify its access control list, before it is ready to serve files.  This ease of use has made such services extremely popular; in 2021, Amazon's S3 service hosted more than 100~trillion files on behalf of its users~\cite{s3trillion}.  However, this same flexibility has given rise to new risks.  The confidentiality of each bucket is not governed by traditional enterprise security mechanisms, but by the correct configuration of individual access control settings by the bucket operator.  Thus, if a bucket is misconfigured to be public, then any party guessing its \emph{name} may gain access to all of its sensitive content.  
In 2018, Ero~\cite{eroBuckets} found that unsolicited parties were indeed guessing the names of public storage buckets.
Shortly after, a survey of almost 200,000 buckets revealed that 10\% contained sensitive data, including passports and financial records~\cite{cable2021stratosphere}.

However, it remains unclear how scanners find such buckets and how many of these are targeted in actual attacks. 
To put it another way: is this merely an abstract risk, or are concrete threat actors actively searching for such vulnerabilities and exploiting them in the wild?
While more than a few high profile breaches have been publicly attributed to misconfigured cloud storage~\cite{studentPII,chinasurv,microsoft,inmates,seniors,capita}, none have documented how these attacks took place, nor the mechanism by which attackers identified the opportunity.   
Indeed, scanning for buckets is non-trivial as an attacker must correctly guess the bucket's full name; the potential search space of these names is $10^{62}$ times larger than IPv6 and no public repository of buckets-in-use exists. The names of misconfigured buckets must therefore either be guessed, or found in an unrelated passive data source (e.g., DNS).

In this work, we empirically analyze this question by deploying a range of ``honeybuckets'' on the AWS S3 platform, configured with names, permissions, and content to lure and measure different scanning strategies.  
By modulating how our buckets are named and whether they are leaked to other data sources, we have been able to identify the most widely used strategies employed by third-parties to scan for misconfigured buckets.  We identify that there are clear preferences for scanning particular kinds of organizations, notably commercial entities in the technology sector.  Moreover, while we find that it is common for all such actors to hide behind proxy servers, we show how to automatically group seemingly disparate IP addresses by dynamically modulating filename content to create causal dependencies between metadata and attempts to access individual files. 

To distinguish between potential benign actors who may be scanning to help notify vulnerable parties~\cite{grayhat} (or at least to try to sell a subscription to such a security scanning service) and those who have malicious aims, we further configured our honeybuckets to create multiple opportunities for actors to engage in clearly malicious acts.  First, we configure our buckets to allow actors to delete or upload data.  We consider users who delete data or upload content that is designed to gain unauthorized access (i.e., to compromise a user who interacts with it and spawn a reverse shell) to be malicious.  Second, we create lures whose value requires one to affirmatively violate a security norm (i.e., an unauthorized login to a third-party server exploiting an ssh credential extracted from one of our buckets). We show that all of these behaviors occur in our data.

In summary, we provide strong empirical evidence that shows how unsecured data on cloud storage is exploited today.
While many targeted scans may reflect benign security interests, there is a range of malicious activity targeting commercial data. We conclude with a range of recommendations for how organizations might better protect such cloud assets in practice.   

%% file: related.tex
\section{Background and Related Works}

Cloud storage mimics a traditional file system interface.
Files are stored in a file directory structure, with the top level directory referred to as a ``bucket.''
Buckets are simple to create: a client must, at minimum, 
(1) choose a service-wide globally unique name that is 3--64 alphanumeric-symbolic characters long and 
(2) configure the bucket contents to be private (the default option) or publicly accessible. 
Once a bucket is created, it is accessible via the cloud provider's API, cloud browser interface, or through a cloud-specific subdomain (e.g., mybucket.s3.amazonaws.com).
A user with sufficient access can upload an unlimited number of files and delete files, among other bucket operations~\cite{aws_s3_operations}.

Once a bucket is accessible (i.e., its access controls allow public access), a third-party must know its \emph{name} to access it.  There is no mechanism for enumerating  names, nor any public repository of bucket names (public or otherwise).  Thus, any party who does not possess \emph{a priori} knowledge of a bucket's name must either guess it -- from a potential namespace of roughly $10^{101}$ possibilities -- or find it in another data source.


An attacker seeking to narrow the search space for guessing such names----to find and sell stolen data, for example~\cite{chinasurv}---can use one of several scanning methodologies.
The simplest bucket scanners rely on a pre-defined list of strings and patterns (e.g., Slurp~\cite{slurp}, s3enum~\cite{s3enum}, and BucketStream~\cite{pastebin-bucketTGA}), tailored towards targeting likely-popular bucket names.  Similar to password guessing software, bucket scanners generate target names by mechanically combining dictionary words, technology terms, etc. 

However, not all bucket names will be simple or use popular words---for example,
a number of applications generate random bucket names (e.g., project ids) for intermediate storage~\cite{appspot}.  
Continella \emph{et al.}\ show that some such names can be identified using passive DNS collection services~\cite{continella2018there} (i.e., where another party exposes the name via their DNS lookups) and Cable \emph{et al.}\ generalize this idea by showing that such data can be used to train a machine learning model to generate a large set of valid names~\cite{cable2021stratosphere}.
\looseness=-1

Such approaches have been used by the research community to explore the storage bucket ecosystem and empirically establish the widespread existence of misconfigured buckets.  Indeed, both   
Continella \emph{et al.}~\cite{continella2018there} and Cable \emph{et al.}~\cite{cable2021stratosphere} found thousands of public buckets that exposed sensitive data, including private keys and national defense documents.  Ero~\cite{eroBuckets} and Cable \emph{et al.}\ further deployed empty storage buckets to establish the existence of online bucket scanning behavior.  
However, while both found evidence of unsolicited scans, neither investigated the method by which names were targeted, nor the actions taken once such buckets were found. 
We are the first to demonstrate how public storage buckets are attacked (i.e., when actors delete, modify, upload, and exploit content) in the wild.

Overall, attacks on cloud storage have been relatively understudied, especially when compared to the broad literature characterizing active attacks on other infrastructure namespaces such as the IPv4 address space~\cite{mirai,bock2021weaponizing,lzr}, IPv6 address space~\cite{barrera2011back,perta2015glance,ullrich2014ipv6}, cloud compute services~\cite{izhikevich2018building,heyYou,yelam2021coresident,izhikevich2023cloud}, DNS~\cite{akiwate2022retroactive,randall2021home,sommese2022observable}, and BGP~\cite{cho2019bgp,demchak2018china,gavrichenkov2015breaking,vervier2015mind}.  In these other environments, researchers have also used honeypots---infrastructure deployed for the purpose of detecting malicious behavior~\cite{kiekintveld2015game}---to characterize IPv4 scanning~\cite{kippo,tpot,cowrie}, email security~\cite{mirian2019hack} and DNS activity~\cite{oberheide2007characterizing}.  Our work brings this approach to the cloud storage context, particularly in service to understanding more about the nature of unsolicited visitors---how they target victims and the extent to which they reveal clear malicious intent.

%% file: phase1.tex
\section{Pilot study: How Buckets Are Targeted}
\label{sec:phase1}

This section describes our pilot\footnote{We use the results of the pilot study to inform a refined experiment in Section~\ref{sec:phase2}.} experiment to broadly understand how, and to what extent, buckets are targeted for attack.
Using a deployment of more than one hundred honeybuckets over a period of six months, we
show that exploitation of misconfigured buckets is very real: hundreds of IP addresses attempted to download, delete, or upload objects---including malicious shell scripts. 
Notably, the first honeybucket was scanned only
40~minutes after deployment.  
Buckets named after companies, universities, and
government organizations were scanned and attacked the most, implying that actors intentionally
scan for specific targets.

\subsection{Methodology}
\label{sub:sec:phase1Meth}
To understand how actors scan and interact with sensitive bucket content, we deployed 112 unique buckets (i.e., ``honeybuckets'') on the AWS S3 platform on February 18, 2022, and hosted them for 6 continuous months. 
We configured the bucket names, permissions, and contents to attract bucket-scanning actors that use various scanning strategies to find public buckets.
We focused only on the AWS S3 platform, as prior work found that AWS S3 receives the largest amount of unsolicited scanning traffic~\cite{cable2021stratosphere}. We do not use AWS decoy resources~\cite{aws_decoy}, as these decoys neither provide bucket naming schemes nor sample content. 

\subsubsection{Bucket Names}
\label{sub:sub:sec:bucketnames_p1}
Scanning buckets is not a trivial task. 
The enormous search-space of bucket names 
is too large to exhaustively enumerate. 
Thus, actors must employ directed strategies for scanning for public buckets.
We used five approaches to name buckets, each designed to measure a unique bucket-scanning strategy.
We list all bucket names in the Appendix in Table~\ref{tab:p1bnames}.

\vspace{3pt}
\noindent
\textbf{Target Generated Buckets.} 
\label{sub:sub:sec:TGA}
To increase the success rate of finding public buckets, actors can use an open-source tool to discover commonly-named S3 buckets. 
These tools generate their scanning targets using target generation algorithms (TGAs), a method that uses a pre-defined list of strings and patterns---often common bucket names---and optionally concatenates them with a user-provided list of keywords. 

To measure if actors use open-source tools to find buckets, we named 12 of the 112 buckets with names generated by three of the most popular bucket enumeration tools found on Github and Pastebin---Slurp~\cite{slurp}, DNSpop~\cite{dnspop}, and bucket-stream-permutation-feature~\cite{pastebin-bucketTGA}.\footnote{We also considered s3enum~\cite{s3enum} and s3Mining~\cite{s3-mining}, but after comparing outputs we found that their generated bucket names were proper subsets of the other TGAs. 
}
To identify an actor's use of a particular enumeration tool, we computed the disjoint set of target names that belong to each tool---and not to any other---to create a set of honeybucket names that are likely to be found by one---and only one---tool.
We chose four unique bucket names from each disjoint set, for a total of 12 buckets covering the three tools.
To act as a control group, we created four additional buckets named with strings that did not appear in any of the bucket enumeration tools.
Crucially, all chosen target names were a part of a fixed set of names that the TGA scans, no matter if the user provides the keyword; it is not clear if this is a bug or a feature of the TGAs.

\vspace{3pt}
\noindent
\textbf{Company, University, and Government Buckets.}\quad
\label{sub:sub:sec:orgs}
Rather than only scanning for popular bucket names, an actor might curate a list of names that target a specific entity.
To detect whether actors explicitly search for buckets named after organizations,
we named 48 of the 112 buckets after the names of companies (Tesla, Walmart, Tinder), government organizations (FBI, CIA, NYPD) and universities (UCSD and Stanford).
We concatenated each of the eight organizations with six unique keywords that appeared in a subset of the bucket enumeration tools.
The six unique keywords consisted of three sensitive (``hidden'', ``private'', ``security'') and three non-sensitive (``production'', ``download'', ``public'') keywords, to additionally measure if a sensitive bucket name influences the type of organizational bucket scanners search for.

\vspace{3pt}
\noindent
\textbf{Cryptocurrency Buckets.}\quad 
If an actor targets specific content, rather than a specific entity, they might scan for buckets named after the content. 
To detect actors who may be searching for buckets storing cryptocurrency, we named four of the 112 buckets after two cryptocurrencies: Bitcoin and Ethereum. 
At the time of our experiment, many variations of ``bitcoin'' and ``ethereum'' bucket names already existed, so we hyphenated the names of the cryptocurrencies with keywords found in the most popular bucket enumeration tools (e.g.,``bitcoin-confidential'').

\vspace{3pt}
\noindent
\textbf{Sensitive and Non-Sensitive Buckets.}\quad 
To search for sensitive content, but not restrict oneself to a specific entity or content-type, an attacker might scan for bucket names with sensitive keywords. 
To compare the discovery rate of buckets named after sensitive keywords (e.g.,  ``passport'' and  ``bank'') to non-sensitive keywords (e.g., ``pictures'' and ``pretty''), we named four of the 112 buckets using each of these keywords.\footnote{We used sensitive keywords defined by the U.S. Justice Department~\cite{sensitive_keywords} and non-sensitive keywords defined by the disjoint set of those sensitive keywords and the top 1000 most common English words~\cite{1000common}.}
At the time of our experiment, bucket names with only the keyword already existed, so we hyphenated all sensitive and non-sensitive keywords with a non-sensitive keyword found in the most popular bucket enumeration tools (``10'').
\looseness=-1

\vspace{3pt}
\noindent
\textbf{Leaked-Alphanumeric Buckets.}\quad 
\input{tables/bucketcontents}
Rather than blindly guessing names, an actor might harvest bucket names from client activity (e.g., DNS queries) to increase the likelihood of discovering buckets. To measure if scanning actors are harvesting names,
we assigned 40 out of the 112 honeybuckets with ``unlikely-guessable'' names: randomly generated alphanumeric names of length 16 (e.g., ``q81osr2ba5wnid4g"). To identify potential sources of leaked buckets,
we leaked 20 of our 40 unlikely-guessable honeybuckets across a variety of platforms, including a Github repository, a new Pastebin repository, a single tweet on a new Twitter account, and the HTML of an academic website (but not visible in a browser). 
We leaked two buckets at a time on each platform to verify whether a scanner likely guessed the bucket by chance (i.e., visited only one bucket) or likely found the bucket on the leaked platform (i.e., visited both buckets).
To identify if scanners used passive DNS as sources for bucket names, we also queried the domains of two unlikely-guessable honeybuckets across DNS resolvers: two resolvers operated by Google, two by Spectrum, two by AT\&T, two by a Russian ISP (ASN\,12714), and two by a Chinese ISP (ASN\,4134). 
Finally, to identify if scanners, such as Google bots, 
unsolicitedly download content from bucket names found in email, we saved two unlikely-guessable honeybuckets in a single email draft on Gmail.
We withheld and did not leak the remaining 20 of 40 alphanumeric buckets to serve as a control group.

\subsubsection{Bucket Permissions}
Having chosen candidate bucket names to scan, an attacker could use one or more of Amazon's 100 operations to interact with the bucket.
To capture as many potential interactions as possible,
we gave bucket-scanning actors considerable freedom to interact with our honeybuckets: all honeybuckets allowed any actor to read the contents of the bucket and write new content to the bucket. However, deleting files originally uploaded by us was forbidden (although we could detect deletion attempts).
We enabled bucket versioning---a feature that saves all past versions of files in a bucket---to track how actors upload, delete, and modify the files they themselves upload.
We recorded all available metadata of interactions with the honeybuckets, including the time of interaction, actor's IP address, actor's AWS account (if the actor sent an authenticated request~\cite{auth2not}), request URI, and any error messages the actor received if their request was malformed or forbidden.
To promote reproducibility, we will share our raw data with researchers. 

\subsubsection{Bucket Contents}
After listing the directory of a bucket, an actor might only choose to download a subset of ``interesting'' files.
We uploaded files with a variety of enticing names and contents to each honeybucket to test for file-download preferences amongst scanning actors. 
Table~\ref{tab:p1contents} lists the nine files we placed in each honeybucket.
To function as controlled variables when comparing download preferences amongst scanning actors, the file names and contents were identical across all buckets.
For example, each honeybucket included a ``Client\_list\_Dec\_2021'' file to lure scanning actors searching for sensitive client information. The file included fake names, home addresses, and social security numbers generated by Faker~\cite{faker}.
Files named ``Backup.pst'', ``Outlook.pst'', ``id\_ed25519'', and ``Inbox.mbox'' were uploaded to lure actors who were searching for sensitive email folder names, SSH private keys, and Google takeout backups, respectively. 
Each file contained fake data in the expected format. 
We additionally included a file with the commonly abused .jar extension~\cite{trojan_jar} to test for malicious actors who might wish to replace existing .jar files with hidden ``trojan'' malware.
The content of the uploaded .jar file emulated a benign calculator program.
Finally, we included two README files---sized 0 bytes and 2 kilobytes---to test if actors checked for file size prior to downloading.

\subsection{Pilot Study Results} 
\label{phase1Res}
In this section we characterize how scanning actors engaged with our honeybuckets.
We investigate the most common methods actors used to scan for buckets, the type of abusive activities buckets received, the amount of time actors spent
interacting with a public bucket, and who was hunting for buckets.
Most notably, we found that \textbf{buckets named after companies were the most likely to be accessed} (Section~\ref{sub:sub:sec:buckets_found}).
Although the majority of bucket interactions only checked for bucket existence, \textbf{hundreds of IP addresses attempted to download, delete, or upload objects}---including malicious shell scripts (Section~\ref{sub:sub:sec:interactions}).
This activity happened quickly after the buckets become accessible: \textbf{actors scanned public buckets within 40~minutes of deployment and uploaded unsolicited content within 10~days} (Section~\ref{sub:sub:sec:time_to_compromise}).

\subsubsection{How Buckets Were Found}
\label{sub:sub:sec:buckets_found}
\input{tables/totalal_num_ops}
\input{tables/ipasperb}
Buckets named after companies were scanned with the most number of operations and IP addresses. 
Table~\ref{tab:topbuckets} lists the top~20 buckets with the most attempted operations. 
Half of these top~20 buckets had a company in their name, with the top five named after ``Tesla.''
For a detailed breakdown,
Table~\ref{tab:ipas} presents the number of unique IP addresses and Autonomous Systems (ASes) that targeted each bucket on average per bucket type. 
Company buckets were targeted with statistically significantly\footnote{We used a one-sided Mann-Whitney U test to evaluate whether the volume of traffic per day that targeted a specific bucket type was stochasticlly greater than the volume that targeted the bucket type with the next greatest volume of traffic per day. We used $p<0.05$ and additionally applied a Bonferroni correction---to account for multiple comparisons---when determining statistical significance.} more IPs and ASes per day compared to all other bucket types.
On average, at least one unique IP and AS visited a company bucket per day.
In Section~\ref{sec:phase2}, we further investigate why actors are lured towards particular companies by deploying a second honeybucket experiment. 

Seven of the top 20~buckets with the most number of interactions had names from open source bucket target generators (i.e., TGAs); however, the majority of actors did not appear to use only the TGAs to generate these bucket names.
Rather, actors were likely using their own list of target bucket names that coincided with the TGA's list. 
We considered an actor to be using a TGA if (1) a single IP address\footnote{The majority of actors used a single IP address when scanning (Section~\ref{sec:phase2}).} targeted all four bucket names that belonged to that TGA; or (2) all TGA buckets were scanned by a uniform distribution of unique IP addresses---which accounts for actors that used multiple IPs (e.g., VPNs) when scanning. 
In Table~\ref{tab:ipas}, we filtered for IP addresses that targeted all four buckets from a single TGA and found statistically significantly fewer actors that exhaustively used the TGA names compared to the number of unique IPs that scanned buckets named after organizations.
Furthermore, we found TGA buckets were not targeted by a uniform number of IP addresses (e.g., ``origin-www'' from DNSpop was targeted by 349 IPs, whereas ``lyncdiscover'' from DNSpop was targeted by 278 IPs).

\input{tables/comp_compare}
Buckets named after universities or a government service were the second most likely to be scanned, with no statistically-significant difference between the two (Table~\ref{tab:ipas}). Among the buckets associated with organizations, bucket names concatenated with the word ``production'' were scanned the most.
Table~\ref{tab:compcompare} shows the relationship between the number of scanning IPs and the organization/keywords in the bucket name.
While 467 unique IPs scanned ``teslaproduction,'' only 375 unique IPs scanned ``tesladownload.''
On the other hand, across all organization types, bucket names concatenated with the word ``hidden'' were an order of magnitude less likely to be targeted than all other keywords  (e.g., 4 compared to 127 unique IPs for ``fbihidden'' vs ``fbisecurity'').

Buckets leaked to passive data sources were the least likely to be scanned: an average of just 0.46 unique IP addresses visited a leaked bucket per day compared to an average of 1.63 IPs that visited a company bucket per day. 
Since the search space for buckets is vast, prior work has found that targeting shorter and lower entropy names results in an overall higher hit rate when scanning buckets~\cite{cable2021stratosphere}.
However, our results imply that actors are optimizing to find specific targets in addition to maximizing overall hit rate.
For example, while bucket ``612'' contained the least amount of entropy in our honeybucket set, it was scanned by fewer unique IP addresses than five of the six Tesla buckets (Table~\ref{tab:topbuckets}), each of which was substantially longer and higher in entropy. 
Using the  Kolmogorov-Smirnov test, we found no statistically significant difference between the number of unique IPs and ASes that targeted buckets ``612'' and ``teslaproduction'' (the Tesla bucket with the highest number of operations) per day, indicating that maximizing overall hit rate was not the only popular scanning strategy. 
Overall, we found that buckets named after companies were scanned the most compared to TGA buckets, leaked buckets, and buckets with names of a lower entropy. 

\subsubsection{Bucket Interactions and Abuse} 
\label{sub:sub:sec:interactions}

Over 100 unique IP addresses (0.9\% of all IPs) uploaded at least one file to a bucket, for a total of 206 files. 
Through manual investigation, we identified four unique files that hosted malicious content (uploaded across 16~unique buckets): 
(1) a ``poc.jsp/'' JavaScript file that spawned a reverse shell to a server specified by a command-line argument; 
(2) a ``test-file.svg'' file that, when opened, re-directed to a suspicious domain (``ngrok.io''); and 
(3) two files, named ``\_snapshot/test'' and ``\_snapshot/test2'', that contained code to send the contents of the \texttt{/etc/passwd} file to the actor who uploaded the file.  

Only two files, ``upload.png'' and ``s3sec.txt,'' contained a message to the bucket owner as a warning that their bucket was public. 
Notably ironic, these good-samaritan warnings were shared through an unsolicited upload. 
Of the remaining uploaded files, six unique files uploaded across nine buckets contained benign content, six files were not accessible due to the object permissions set by the actor, and 188 files were empty.
We summarize the contents of all uploaded files in Table~\ref{table:uploaded_files} in the Appendix.

Over 700 unique IP addresses (6\% of all IPs) attempted to download a file from a unique bucket. 
The client list file was downloaded the most, with 160 total downloads across 88 unique IP addresses  (Table~\ref{tab:p1contents}). 
We did not find any file to have a statistically significantly\footnote{We calculated statistical significance using the methodology from Section~\ref{sub:sub:sec:buckets_found}.} greater number of downloads per day, likely due to the infrequent nature of downloads per day (i.e., on average, there are 0.00005 downloads per bucket per day). 

We summarize the remaining non-abusive bucket interactions in Appendix~\ref{app:exp1}, which often consists of checking a bucket's existence and/or listing files. 

\subsubsection{Time-to-Compromise of Buckets}
\label{sub:sub:sec:time_to_compromise}

Actors were quick to find buckets:
within 40 minutes, the first bucket---``walmartdownload''---had its directory listed. 
However, it took at least one week for an actor to abuse a bucket (i.e., upload an unsolicited object to the ``lyncdiscover'' bucket generated with the TGA DNSpop). 
We considered a bucket to be abused when an actor attempted an upload, download, or delete operation on the bucket. 

Buckets named after TGA targets or company names were the only two categories to experience uploads, with a bucket receiving a first upload within an average of 71~days. 
The first successful upload across all 112~buckets occurred 10 days after deployment in which an unauthenticated actor uploaded a file with instructions on how to make the bucket private.
The file was called ``s3sec.txt'' and can be found in the Github repository s3sec~\cite{s3secgithub}. 

Compared to uploading content, actors were much slower to download content.
The average time-to-first download across all bucket types was 78 days, with the first successful download of the file ``Backup.pst,'' 27 days after deployment, in the bucket ``tesladownload.''
Content inside the non-sensitive keyword and control buckets was never downloaded over the course of the experiment.

The most rare and slowest-to-occur abuse of a bucket was file deletion: the only attempt to delete a file from Table~\ref{tab:p1contents}

occurred 134~days after bucket deployment. Thus we conclude that actors were quick to find buckets and overall performed abusive operations by uploading unsolicited files and downloading files, but rarely attempted to delete our files from our buckets.

\subsubsection{Identifying Bucket-Scanning Actors}
\label{sub:sub:sec:auth_actors}
\input{tables/auth2ASLeft}

A total of 6,567 unique IP addresses performed at least one operation to at least one bucket.

Nearly all (99.9\%) IP addresses sent an unauthenticated AWS request,\footnote{Amazon allows users to be unauthenticated, which is when a user does not have an AWS account~\cite{auth2not} or when an authenticated user adds the flag ``--no-sign-request'' to their command line argument~\cite{unauthyourself}.} allowing the user to remain anonymous. 
However, 27 actors authenticated themselves, and eight used non-alphanumeric usernames.
Table~\ref{tab:auth2as} lists the eight actors, providing their username, IP addresses used, buckets visited, etc. 
Three actors had usernames that alluded to bugfinding (i.e., ``s3bug'', ``bug'', ``pudsec'').
The user ``bug'' uploaded a ``Read.txt'' file that described how to pen-test buckets for the purposes of receiving a bug bounty.
Three remaining users used administrator accounts, in which at least one alluded to being a ``bot'' for scanning (i.e., ``Admin.../xbotusr''). 
The remaining 19~authenticated users used non-informative, random alphanumeric names. In Section~\ref{sec:phase2}, we deploy a new set of experiments to filter for only non-bot scanners.

Authenticated users also gave a glimpse into understanding if bucket-scanning actors often used multiple source IP addresses, and whether a unique IP address was likely to identify a unique scanning actor.
We used the authenticated actor set as an approximate ground truth mapping of unique users to IP addresses.\footnote{The set is likely biased towards users that did not care about concealing their identity and thus served as an expected upper-bound of the number of actors that used only one scanning IP address.}
IP addresses were a sufficient approximate indicator of unique scanning actors: 90\% of authenticated actors used only one scanning IP address and 100\% of authenticated actors used IP addresses from the same autonomous system (Table~\ref{tab:auth2as}). 
In Section~\ref{phase2Meth}, we show that the vast majority of bucket scanning actors were likely only using one IP address.

All scanners originated from a set of 330 autonomous systems, a subset of which have security-critical reputations. 

Approximately 66\% of scanners originated from three ASes: M247 (ASN\,9009), HostRoyale (ASN\,203020), and CHOOPA (ASN\,20473).
Clients using M247 and Choopa are known to be consistently engaged in high-risk and highly fraudulent behavior~\cite{m247_fraud,choopa_fraud}.
HostRoyale's clients are known to use its anonymizing VPN services~\cite{hostRoyale_fraud}.

\subsection{Summary}

This pilot study systematically demonstrated that 
scanning strategies are not random: our honeybuckets named after well-known
companies received the most activity (Section~\ref{sub:sub:sec:buckets_found}). Scanners were quick
to discover new publicly accessible buckets, finding
buckets within 40~minutes of deployment (Section~\ref{sub:sub:sec:time_to_compromise}). Finally, scanning actors
actively interacted with the bucket contents in a variety of concerning
ways: at least one file containing sensitive data was downloaded
across nearly all buckets, many buckets had malicious files uploaded
to them, and some had files targeted for deletion (Section~\ref{sub:sub:sec:interactions}).

While this experiment established many aspects of scanning activity, a
number of questions remain.  It is unclear why actors were lured
towards particular companies, how actors might take further
advantage of downloaded sensitive data, and whether identical files across buckets could cause actors to recognize the decoys and modify their behavior.  
Furthermore, actors can use
multiple IPs and VPNs to mask their activity, and such aliasing leaves
unresolved how to attribute multiple interactions to the same actor.
In the next section, we build on our pilot study and deploy a new, refined honeybucket experiment to provide more insight into precisely these questions.

%% file: tables/bucketcontents.tex
\begin{table}[t]
\caption{\textbf{Honeybucket File Contents}---\textnormal{Each honeybucket hosted nine unique files that contained a variety of sensitive names and content, intended to attract scanning actors with different target preferences.}}
\begin{tabular}{llrrr}
\toprule
 File name            & Content                         &   Unique &   Unique &  Down- \\
                      &                                 &  IPs &  ASNs & loads\\
\midrule
 Client\_list\_ & Fake  &           88 &            56 &                 160 \\
Dec\_2021                         &  names, &              &                &  \\
&SSNs,&&&\\
& addresses& & & \\
 \midrule
 Backup.pst           & Sensitive &           69 &            47 &                 155 \\
 &mail&&&\\
  &folder&&&\\
                      &names&               &               &               \\
 \midrule
 README1              & AAA...&           64 &            41 &                 127 \\
 \midrule
 Outlook.pst          & Sensitive&           54 &            34 &   110 \\
                       &mail&               &               &               \\
                        &folder&&&\\
                       & names &&&\\
 \midrule
 README2              & Empty &           53 &            32 &                 112 \\
 & File&&&\\
 \midrule
 id\_ed25519           & SSH &           53 &            31 &                 117 \\
 &private&&&\\
 &key&&&\\
 \midrule
 Inbox.mbox           & Google &           50 &            32 &                 107 \\
 &Takeout& & & \\
 & backup & & & \\
 \midrule
 UTC...               & UTC &           48 &            30 &                 112 \\
 &Wallet&&&\\
 \midrule
 javazoom.jar         & Benign  &           41 &            21 &                 103 \\
 &jar&&&\\
 &file&&&\\
\bottomrule
\end{tabular}
\vspace{8pt}
\label{tab:p1contents}
\vspace{-5pt}
\end{table}

%% file: tables/totalal_num_ops.tex
\begin{table}[t]
\caption{\textbf{Top 20 Buckets With The Most Attempted Operations}---%
\textnormal{Five out of six buckets named after Tesla experienced the most attempted operations.
}}
\centering
\begin{tabular}{ccrr}
\toprule
 Bucket name                   & Type  & \# Ops & \# IPs\\
\midrule
 teslaproduction       & Company       & 2538  & 467\\
 \midrule
 teslapublic           & Company          & 1620  & 355\\
 \midrule
 tesladownload       & Company    & 1601   & 375\\
 \midrule
 teslasecurity         & Company     &  1470  & 339\\
 \midrule
 teslaprivate   & Company   &  1456   & 342\\
 \midrule
 origin-www & TGA  &  1379     & 349\\
 &(DNSpop)&&\\
 \midrule
 612       & TGA  & 1312    & 446\\
  &(DNSpop)&&\\
  \midrule
 lyncdiscover  & TGA        &  980    & 278\\
  &(DNSpop)&&\\
  \midrule
 www-download & TGA & 894  & 323\\
  &(Pastebin)&&\\
  \midrule
 walmartproduction & Company & 872  & 170\\
 \midrule
 tinderproduction & Company & 755  & 178\\
 \midrule
  ucsdprivate & Universities & 747  & 143\\
 \midrule
 fbiproduction & Government & 627  & 179\\
 \midrule
 www-slack & TGA & 604  & 260\\
   &(Pastebin)&&\\
   \midrule
 www-security & TGA & 572  & 220\\
   &(Pastebin)&&\\
   \midrule
 screenshots-www & TGA & 552  & 245\\
   &(Pastebin)&&\\
   \midrule
 tinderpublic & Company & 542  & 139\\
 \midrule
 tinderdownload & Company& 527  & 129\\
 \midrule
 walmartsecurity & Company & 516  & 133\\
 \midrule
 ciaproduction & Government & 508  & 151\\
\bottomrule
\end{tabular}
\vspace{8pt}
\label{tab:topbuckets}
\vspace{-15pt}
\end{table}

%% file: tables/ipasperb.tex
\begin{table}[t]
\caption{\textbf{Traffic Across Bucket-Types}---%
\textnormal{Buckets named after companies experienced the most traffic on average both overall and per day. 
Statistically significant increases of traffic per day, relative to all other types buckets that experienced less average traffic per day, are marked with a \texttt{*}.
For example, while university buckets did not experience statistically significantly more traffic than government buckets, government buckets did experience significantly more traffic than all bucket types except for company buckets.
}}
\centering
\begin{tabular}{cS[table-format=3.2]S[table-format=2.2]S[table-format=1.2]S[table-format=1.2]}
\toprule
 Type     &   IPs  & ASNs  &   IPs   &  ASNs \\
   \cmidrule(r){2-5}
   &\multicolumn{4}{c}{\# unique on average per bucket}\\ 
   
\cmidrule(r){2-3}\cmidrule(r){4-5}
     & \multicolumn{2}{c}{total} &  \multicolumn{2}{c}{per day}  \\

\midrule
  Companies              & 195.39   & 35.72 &  1.63*     &  1.44* \\
  \midrule
 Universities           & 133.83   & 20.75 &  1.27     &   1.19 \\ 
 \midrule
 Government             &  100.22   & 19.5  &   1.08*  &   1.04* \\
 \midrule
 Non-sensitive  &  74.50    & 15.00 & 0.95&    0.90  \\
 Keywords &&&&\\
 \midrule
  Sensitive    &  43.50    & 8.50   & 0.74     &  0.72 \\
  Keywords &&&&\\
  \midrule
  Cryptocurrency         & 27.75   & 7.75 & 0.56*     & 0.54* \\
  \midrule
   TGA     & 12.67 &  12.67  & 0.42* &   0.46* \\
   - filtered &&&&\\
   \midrule
 Leaked                 &  6.17   & 3.22  & 0.46*     &  0.45 \\
 \midrule
 Control             & 1.17   & 1.17 & 0.44&  0.44* \\

\bottomrule
\end{tabular}
\vspace{8pt}
\label{tab:ipas}
\end{table}

%% file: tables/comp_compare.tex
\begin{table*}[t]
\caption{\textbf{The Impact of Bucket Name Construction on Received Scans}---%
\textnormal{Buckets that contained the keyword ``production'' were targeted by more unique IP addresses than buckets containing any other keyword, no matter the organization.}}
\centering
\begin{tabular}{lrrrrrr|r}
\toprule
  \multicolumn{1}{c}{Name} &\multicolumn{7}{c}{\# Unique IPs Targeting Org-Keyword Bucket}\\ 
  \cmidrule(r){1-1}\cmidrule(r){2-8}
   &   production &   download &   public &   private &   security &   hidden &   Total \\
    
\midrule
 Tesla                &          467 &        375 &      355 &       342 &        339 &        8 &    1886 \\
 Walmart              &          170 &        142 &      140 &       139 &        133 &        8 &     732 \\
 Tinder               &          178 &        129 &      139 &       125 &        108 &        5 &     684 \\
  \midrule
Public                &          130 &        131 &      128 &       143 &        129 &        3 &     664 \\
University & & & & & & & \\
Private           &          163 &        173 &      162 &       158 &        158 &        3 &     817 \\
University & & & & & & & \\
  \midrule
 FBI                  &          179 &        122 &      119 &       116 &        127 &        4 &     667 \\
 CIA                  &          151 &        116 &      116 &       117 &        118 &        4 &     622 \\
 NYPD                 &           78 &         78 &       76 &        78 &         78 &        3 &     391 \\
 \midrule
 Total                &         1516 &       1266 &     1235 &      1218 &       1190 &       38 &     6463 \\
\bottomrule
\end{tabular}
\label{tab:compcompare}
\end{table*}

%% file: tables/auth2ASLeft.tex
\begin{table*}[t]
\caption{\textbf{Authenticated User Activity}---
\textnormal{We present a list (8 out of 27) of all authenticated users who used non-alphanumeric usernames (omitting their unique account ID for brevity).
The majority of authenticated users used only one IP address to scan, only visited buckets of one type (e.g., TGA, company), and did not interact with bucket content.}}
\centering
\small
\begin{tabular}{lllll}
\toprule
User   & IPs used  & ASN   & Buckets Visited   & Operations \\
\midrule
 user/energi-0001              & 103.157.116.108/32    &  Cloud Teknologi  (137331)  &   All DNSpop  & Check exist, List dir.,    \\ 
 &&&& Get ACL\\
 \hline
 assumed-role/...             & 12 IPs in  & DC Protection (198949)  & ``origin-www'' (DNSpop) &  Get ACL\\ 
 Admin.../xbotusr&148.177.96/24&& &\\
 \midrule
 user/Admin              & 186.29.129.113/32                 & ETB (19429)    & ``612'' (DNSpop) & List dir., Get ACL \\ 
                         &  190.25.111.135/32               &            & &  \\ %
 \midrule
 user/bref-cli             & 159.89.129.123/32                 & DIGITAL OCEAN (14061)   & ``lyncdiscover'' (DNSpop) & Check exist, List dir., \\ 
  &&&& Get ACL\\
 \midrule
 user/Administrator             & 143.238.166.88/32            & Telstra (1221)    & ``origin-www''  (DNSpop) & Check exist, List dir.,\\ 
 & & &``lyncdiscover''  (DNSpop) &   Get ACL\\
 \midrule
 user/s3bug             & 103.105.154.178/32               &   Global Ra Net (135692)   & 3/4 Tinder buckets  & Get ACL\\ 
 \midrule
 user/bug             & 103.79.171.204/32               & MNR Broadband (133648)  & ``tinderpublic''  & List dir., Upload object \\ 
 \midrule
 user/pudsec             & 216.126.238.240/32           & Hostodo (399804) & ``612'' (DNSpop)  & Check exist, List dir., \\ 
  &&&& Get ACL\\
 \bottomrule
\end{tabular}
\vspace{8pt}
\label{tab:auth2as}
\end{table*}

%% file: phase2.tex
\section{Exploitation of Company Buckets}
\label{sec:phase2}

In our pilot study, actors were most likely to scan and download sensitive files from buckets named after companies, while occasionally using multiple IPs. 
In this section, we conduct a new, refined experiment to broadly investigate (1)~what types (e.g., industry sector, Fortune 500 standing, having a vulnerability disclosure program) of companies receive the most traffic, (2)~if actors using multiple IP addresses can be more easily identified and, most importantly, (3)~whether downloaded content is exploited. 
We indeed found that companies with a vulnerability disclosure program were more likely to be scanned.
Most alarmingly, we recorded eight instances of actors exploiting downloaded content from our buckets, which directly led to unauthorized attempts to login to one of our servers.

\subsection{Methodology for Bucket Configuration}
\label{phase2Meth}
We deployed 120 honeybuckets on the AWS S3 platform on October 2, 2022 and hosted them for 1 month.
We configured buckets with three primary differences from the methodology in Section~\ref{sub:sec:phase1Meth}:
(1) names followed a single enterprise-themed naming scheme, 
(2) buckets contained an informative document that tracked the longevity of information post-download, and 
(3) bucket content used a unique identifier that helped track actors who used multiple IPs.

\subsubsection{Bucket Names} 
\label{sub:sub:sec:bnames}
We investigated what factors cause the buckets of one company to be at a higher risk of abuse (e.g., malicious uploads, malicious downloads) than another company.
To study the cloud-storage attack surface of enterprises, we named 120~new buckets after 
60~Fortune 500 companies~\cite{fort500}.
We created the set of 60~companies by (1) removing company names with an ``\&'', as that symbol is not allowed in bucket names, (2) randomly ordering the remaining Fortune~500 companies, and (3) selecting the first~30 companies that had a clear vulnerability disclosure procedure (VDP) and the first 30 companies that did not appear to have a VDP.
To determine if a company hosted a VDP, we used Google to search for ``$<$company name$>$ vulnerability disclosure'' and looked for a disclosure procedure within the top~10 results.\footnote{This search methodology identified that 21\% of the first set of the 60 randomly chosen companies had a vulnerability disclosure program, which is nearly-identical to what prior work has found~\cite{20VDP}.}
\looseness=-1

To construct the bucket names, we concatenated (with no spaces) each chosen company with the two keywords from Section~\ref{sub:sub:sec:buckets_found} that attracted the greatest number of actors---``production'' and ``download''---thereby assigning two buckets per company (e.g., ``carvanaproduction'' and ``carvanadownload'').
Table~\ref{tab:f500} in the Appendix lists the names of all chosen companies,\footnote{During bucket deployment, we encountered an already-existing bucket, ``Blackrockproduction''. We replaced Blackrock with another randomly-chosen Fortune~500 company.} their offering of a VDP and/or bug bounty, and their 2022 Fortune~500 ranking. In Section~\ref{ethics}, we discuss the ethics of this methodology.

\subsubsection{Bucket Contents}
\label{sub:sub:sec:p2_buc_contents}
Recall that in Section~\ref{phase1Res}, some actors downloaded at least one file from the bucket.
In this experiment, our goals are to (1)~identify what actors do after downloading a file, and (2)~better estimate the number of actors engaging with the buckets. 
To accomplish these goals, all buckets in our second experiment hosted (1)~fictitious sensitive content to lure actors to interact with our honeybuckets in a way that allowed for tracking their actions, and (2)~a new text document that served as a source of information to trace the identity of actors.

\vspace{3pt}
\noindent
\textbf{Sensitive Information.}\quad
To lure actors into interacting with our honeybuckets,
we hosted a nested directory of fake financial data 
generated by the Faker tool~\cite{faker}.
Each honeybucket hosted unique data (unlike our first experiment in Section~\ref{sub:sec:phase1Meth}), to reduce the chance of actors finding multiple company-named honeybuckets and possibly growing suspicious if they encountered identical data. 
We named the nested directory with an hourly-changing hashed time stamp (i.e., ``update\_2022\_chargeback\_\{unix time\}''). As a result, an actor who used multiple IP addresses across multiple hours to list bucket contents and download individual files could be identified using the hashed time stamp. 
To avoid triggering alarms,\footnote{After nine days of deployment, Amazon sent a note to an email address associated with our lab's AWS account, warning us that two out of 120 honeybuckets were ``publicly hosting highly sensitive and confidential information.'' We immediately configured these buckets to be private and renamed 
them to incorporate the names of two new Fortune~500 companies (Table~\ref{tab:f500}) that fit our criteria from Section~\ref{sub:sub:sec:bnames}.} we zip-encrypted the contents of individual sensitive files across all honeybuckets.

\vspace{3pt}
\noindent
\textbf{Informative Document.}\quad
All buckets hosted a single, unencrypted text document that (1) encouraged actors to contact the bucket owner, (2) traced the longevity of sensitive information post-download, and (3) identified actors who used multiple IP addresses. 
To encourage actors to contact us, the document included an email address that we controlled, and attributed bucket ownership to an ambiguous financial analytics contractor of the Fortune~500 company. 
In this way, we could still measure interactions associated with buckets named after Fortune~500 companies, but---assuming the actor read the informative document---re-direct follow-up email interactions to us. 

We also wanted to infer whether actors who downloaded sensitive data had malicious intentions, such as using the downloaded information for nefarious purposes.
For this goal, in the document we also included an SSH username, password, and IP address for a Cowrie SSH honeypot~\cite{cowrie} that we hosted.
To lure an actor to login via SSH, we (falsely) stated that the encryption key to the sensitive files in the bucket could be found in our SSH honeypot. 
While no username or password combination granted entry into our Cowrie honeypot, we monitored login attempts (i.e., IP address, timestamp, username and password attempted) to see if any attempted SSH credentials matched the credentials provided in the honeybucket. 

In the first experiment (Section~\ref{phase1Res}), thousands of unique IP addresses interacted with our buckets. 
To identify individual actors who may have used multiple IP addresses to search for buckets and SSH into our honeypot, we updated the SSH password in the informative document every hour by concatenating the password with a hash of the current timestamp. 
We then used the SSH password as a link between the IP address that obtained the password and the IP address that used it. 
In a similar manner, we updated the name of the text document to trace actors who may have used different IP addresses to list directory contents and download individual files. 
To attract actors to our informative document, we named the informative document `secure-encryption-ssh-quickstart-\{unix time\}.txt' to attract downloads with its enticing name.
Relative to all other bucket content, the document's name was alphabetically first, which ensured that our informative document appeared first when an actor listed the contents of the bucket. 
We provide the exact text of the document in Appendix~\ref{appSSH}.

\subsubsection{Bucket Permissions}
In Section~\ref{sub:sub:sec:interactions}, we documented actors uploading over 100~files---some of which were inaccessible to us---across different buckets.
To preserve the integrity of our bucket configuration, in this experiment actors were only allowed to (1) list the bucket directory, (2) download all objects, and (3) upload an object if and only if the actor transferred ownership of the uploaded object to the bucket owner. Currently, this is the only solution to automatically ensure uploaded files can be accessed by the bucket owners, since files are automatically owned by the uploader~\cite{aws_files_owner,s3_download}. Deleting objects was forbidden. 

\subsection{Results}
\label{phase2Res}
In this section, we use the increased lures and tracking capabilities to further understand how bucket scanners operate.
We start by investigating what kinds of companies lured the most actors and found \textbf{a significantly increased number of actors and significantly increased amount of abusive behavior correlated with companies with a vulnerability disclosure program} (Section~\ref{sub:sub:sec:understand_comp}). 

We found that certain scanners used VPNs, prompting us to develop a scalable approach to track and group colluding IPs as belonging to the same actor.
Using this approach, we found most actors still only used 1 IP address (Section~\ref{sub:sub:sec:actor_meth}).
Finally, we analyze the ``abusiveness'' of actor behavior, distinguishing between those actors who merely downloaded our purportedly sensitive information and those who used that data to login to another machine that they were not authorized to access. 
\textbf{We traced over 3000 login attempts to 8 unique actors who had previously downloaded content from our bucket} (Section~\ref{sub:sub:sec:ssh_events}).
\subsubsection{Understanding Company Targets}
\input{tables/type_analysis_p2}
\input{tables/top_10}
\label{sub:sub:sec:understand_comp}
Buckets that contained the names of companies with a vulnerability disclosure program (``VDP companies'') were statistically significantly\footnote{We used the same one-sided Mann-Whitney U methodology from Section~\ref{sec:phase1}.} more likely to be scanned.
Table~\ref{tab:typep2} presents the number of IP addresses who scanned different categories of companies. 
On average, VDP-company buckets were scanned by 60\% more IP addresses compared to non-VDP company buckets, attracting roughly 18 IPs per day.
VDP-company buckets were responsible for the majority (6/8) of the abusive SSH behavior (Section~\ref{sub:sub:sec:ssh_events}).
Our results are consistent with previous work~\cite{bountyOutsider} that found a positive correlation between bug bounty programs and data breaches when studying government-reported breaches. One possible explanation is that potential financial incentives attract scanners, but we cannot discount the possibility that companies with valuable online assets are simply more likely to institute VDPs (perhaps due to being attacked more frequently).

Additionally, buckets with names of companies in the technology sector were statistically significantly\footnote{Accounting for Bonferroni correction, the Mann-Whitney U methodology only returns significant p-values for sample sizes greater than 10. Thus, the following company sectors were excluded from statistical analysis: Healthcare (10 buckets total), Transportation (8), Engineering (6), Aerospace (4), and Business Services (2).} more likely to be scanned by unique IPs compared to any other sector.
Buckets named after technology companies\footnote{We determine sector by using the Fortune 500 sector label.} were scanned by 74\% more IPs on average than healthcare buckets, the second-most scanned sector. 
Technology company buckets attracting the most IPs was not a symptom of having a VDP: while 5 out of 8 technology companies had a VDP, 10 out of 10 of the health companies and 10 out of 16 of the financial companies also had a VDP.
We found no correlation between a company's Fortune~500 rank and the number of IP addresses who scanned its corresponding bucket. 
\looseness=-1

Table~\ref{tab:top10} lists the top~ten buckets that were scanned by the greatest number of unique IP addresses.
The two most-scanned buckets were both named after American Express, which has a VDP.
Unique actors consistently scanned American Express buckets throughout our experiment.

Six out of the top ten most-scanned buckets were named after technology companies, and nine were named after companies with a VDP.  
\looseness=-1
\subsubsection{AWS Triggered Reports}
\label{sub:sub:sec:aws_reports}
We received two AWS triggered reports throughout the course of our experiment. These reports were not prompted by an internal AWS hygiene system. Instead, they were instigated by humans (employees or contractors) that explicitly searched for buckets named after these companies. Our first AWS report regarded one of our buckets named after a technology company. AWS included the emails of the ``original abuse reporter[s]:'' two employees of that company. Our second report regarded two of our buckets named after a healthcare company, with the contact information of a security employee from said company. After removing the buckets, as per their request, we spoke with the healthcare security team and learned that the buckets were reported to them by a third-party rather than an internal scanning system or team. Thus, we see that both AWS and the company rely on good Samaritans to report bucket misconfigurations, instead of AWS's internal, automated detection systems~\cite{detective,guardduty}.\footnote{Indeed, to the best of our knowledge, neither the AWS security investigator, Detective~\cite{detective}, nor the AWS threat detection service, GuardDuty~\cite{guardduty}, specifically monitor for buckets hosting exposed sensitive data.} Furthermore, this healthcare company operates a vulnerability disclosure program, reinforcing that the presence of such a program may influence how scanners pick their targets.

\subsubsection{Identifying and Tracking Unique Actors}
\label{sub:sub:sec:actor_meth}
Since some actors are likely to use VPNs to interact with the buckets, we developed an algorithm to identify ``colluding'' IP addresses to better identify and track \textit{at scale} a single actor's operations across multiple IP addresses. We define a single actor using colluding IP addresses to be either one scanner operating under a VPN \textit{or} multiple parties colluding by sharing or selling information to each other.
We used this algorithm to better approximate the true number of actors interacting with the buckets, rather than just using the number of unique IP addresses.\\

\begin{figure*}
    \centering
\includegraphics[width=\linewidth]{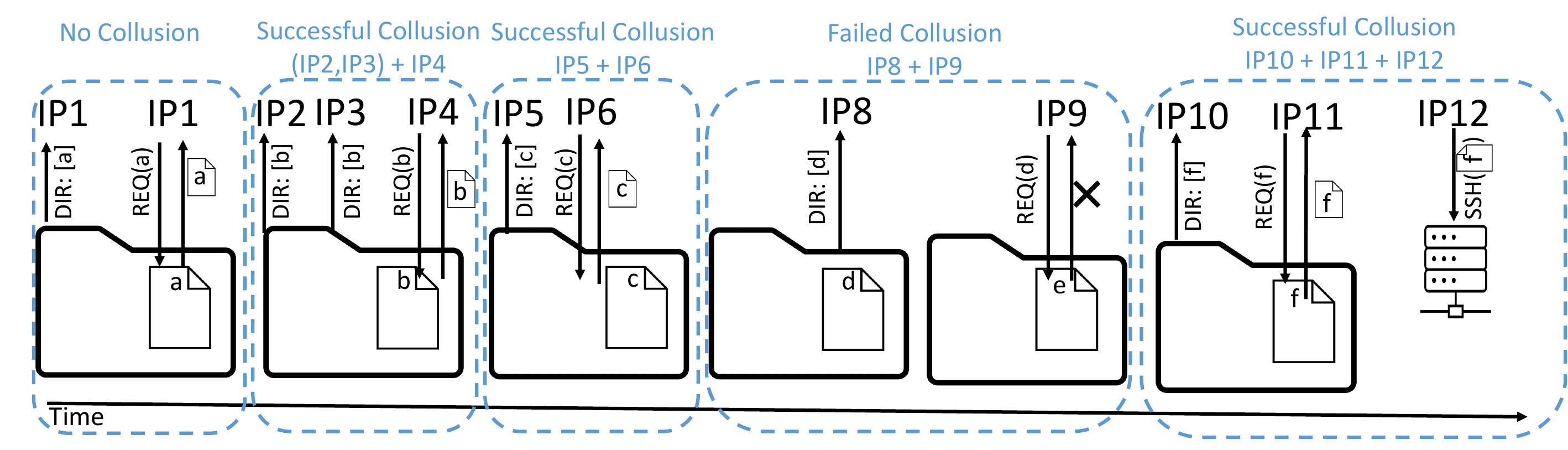}
    \caption{\textbf{Identifying Colluding IP Addresses}---
\textnormal{We identified colluding IPs using the causal dependencies created by hourly-updated filenames. In our figure, ``DIR: [a]'' is the return of a directory listing with the filename ``a,'' ``REQ(a)'' is the request of filename ``a,'' and ``SSH(a)'' is the extraction of SSH credentials from filename ``a.''   
When an IP requested a file without having listed the directory (e.g., IP4), all IPs that listed the directory within the same hour (e.g., IP2 and IP3) were identified as potentially colluding with the file-requesting IP. An IP that SSH'ed into our honeypot server (e.g., IP12), without having downloaded the file with the SSH instructions, was identified as colluding with the IP that downloaded the file with the matching SSH credentials (e.g., IP11).  }}
    \label{fig:SSH_B}
\end{figure*}

\noindent
\textbf{Algorithm.}\quad
To identify colluding IP addresses, we relied on the hourly updates of unique identifiers in bucket filenames (Section~\ref{sub:sub:sec:p2_buc_contents}), which provided a link between colluding IP addresses that execute operations on behalf of each other. 
Concretely, there are three cases of operations that revealed colluding IPs:
\begin{enumerate}[align=left]
    \item[\textbf{Failure:}]
    Without having ever listed the directory, there was an attempt to download a file that used to, but no longer, exists. For example, in Figure~\ref{fig:SSH_B}, IP9 never listed the directory, but attempted to download file ``d,'' which no longer existed. IP8 was the only IP  that listed the directory during file d's existence, thus IP8 and IP9 must have been colluding IPs.
    \item[\textbf{Success:}] Without having ever listed the directory, there was a successful download of a file. For example, in Figure~\ref{fig:SSH_B}, IP6 never listed the directory, but successfully downloaded file ``c''. IP5 was the only IP address that listed the directory after file c was uploaded, but before IP6 downloaded file c, thus IP6 must have been colluding with IP5.
    \looseness=-1
    \item[\textbf{SSH:}] Without having ever downloaded the informative document, there was an attempt to login to the SSH Honeypot. For example, in Figure~\ref{fig:SSH_B}, IP12 never listed the directory, but attempted to login to the SSH Honeypot using the credentials only found in file ``f''. IP11 was the only IP address that successfully downloaded file f, thus IP12 and IP11 must be colluding IPs. We use the Success case above to determine that IP10 and IP11 must have been the same actor as well.
\end{enumerate}

\noindent
We formalize the logic to detect all three cases of likely-colluding events in the Appendix in  Algorithm~\ref{fig:alg_collude}.
Algorithm~\ref{fig:alg_collude} takes as input (1) an IP address
and (2) a token associated with the IP address. The algorithm returns as output a (potentially empty) set of IP addresses that colluded with the input IP address based upon the time of their operations.
The token associated with the IP address
is the informative document or SSH password that the IP address had downloaded or used. The token contains the unique time-based identifier that identifies which time frame another colluding IP could have downloaded the token from.
Upon identifying the time frame, the algorithm simply searches for which other IP addresses downloaded the token in the target time frame, and assigns those IP addresses to the returned colluding set. 
We note that if the input IP address did not download a document or abuse an SSH password (i.e., ``token == none''), then the algorithm will not return any colluding IPs for the current IP. Instead, if it was indeed a colluding IP, it may be returned as a colluding IP for another IP address that did download a document or abuse an SSH password.

Notably, the algorithm's accuracy directly depends upon the granularity of content updates.
For example, since our methodology updates bucket file names every hour, IP addresses that appear together within the same hour---whether by accident or due to collusion---often cannot be differentiated and are all considered candidates for collusion. This is seen with IP2, IP3, and IP4 in Figure~\ref{fig:SSH_B}, where IP2 and IP3 both listed the bucket directory and therefore are both considered candidates for colluding with IP4.
Further, the algorithm cannot detect if two unique actors are using two VPNs within the same hour on the same bucket. 
Nevertheless, as shown in Table~\ref{tab:ipas} the number of unique IPs that visited a bucket per day---let alone per hour---is often too low (i.e., $<$ 1 unique IP per day) for the algorithm's limitation to be a problem. 

\vspace{3pt}
\noindent
\textbf{Characterizing Actors.}\quad
Applying this algorithm across all logged bucket operations, the vast majority (94.6\%) of actors appeared to use only one IP address.
In the most extreme case, one actor used 45~unique IPs---a subset of which map to known VPN products and Tor exit nodes---to download files. 
In Section~\ref{sub:sub:sec:ssh_events}, the actors who used more than one IP address were more likely to participate in security-critical behavior. 
The remaining 5.4\% (61 out of 1128) of IP addresses clustered to at most six unique actors. 

\subsubsection{Abusive access}
\label{sub:sub:sec:ssh_events}
Among the challenges for a study like ours is that threat actors are anonymous and their underlying motivations are undeclared.
Thus, it can be difficult to distinguish between the visits of a benign scanner (e.g., a security researcher or pentester), and a malicious party who is seeking to exploit exposed resources.  It is to address this ambiguity that we introduced the SSH honeypot component of our experiment.  By creating an opportunity for scanners to ``cross the line'' and attempt an unauthorized login to third-party infrastructure using credentials harvested from our buckets, we create a measurement that is clearly interpretable.  Independent of any motivation, such actions are widely understood to violate ethical norms~\cite{zmap} and, in most countries, civil and criminal law as well.~\footnote{For example, in the US unauthorized access of this form is a violation of 18 USC 1030, the Computer Fraud and Abuse Act.}  To put this another way, there is no motivation that excuses an attempt to use purloined credentials to gain unauthorized access to an unknown server. Indeed, it would be particularly worrying if purportedly benign researchers were taking such action. Thus, we consider all such accesses to be malicious.

We tracked the downloads of the informative document and monitored our SSH honeypot over a span of 6 months.  A total of 182~unique IP addresses downloaded at least one bucket's informative document, which contained an email address (i.e., a direct method to contact the bucket owner) and leaked SSH credentials. 
Eight unique IPs, traced to eight unique actors, collectively performed over 3000 login attempts using the leaked SSH credentials.
Only one actor sent an email message.~\footnote{The single message we received was from a party warning us, in our capacity as the purported financial analytics contractor, that our \textit{usaadownload} bucket was public.  This ``good Samaritan'' requested that we pay a bug bounty and sent three follow up-emails insisting on a reward. We note that USAA, which operates a monetary VDP~\cite{usaa}, belongs to the category of organizations that attract the most scanning traffic.}

Using the algorithm outlined in Section~\ref{sub:sub:sec:actor_meth} we have distinguished two categories of SSH-abuse from our data. We call not enumerating all possible passwords ``SSH Attempt,'' which was performed by 5~actors, and call using all 1000 possible passwords ``SSH Brute Force,'' which was performed by 3~actors.

\vspace{3pt}
\noindent
\textbf{SSH Attempt.}\quad
Five unique actors---whose IP addresses belong to Microsoft Cloud (AS\,8075), Charter Communications (AS\,11427), the high-fraud risk Cloudvider (AS\,66240), TIAA (AS\,2923), and NETSPI (AS\,397919)---attempt at most eight passwords against our SSH honeypot. 

In the first SSH case (16~days after bucket deployment), a single actor used two IP addresses to find a company bucket, download the informative document, and attempt to login to the SSH Honeypot---all within 3~minutes.
The actor used address IP1 to list the directory of the bucket ``tiaadownload'' and download the informative document. Within two minutes, address IP2 attempted to login to the SSH Honeypot using both the username and password originally displayed to IP1 a few minutes prior. IP1 was the only address to download the informative document with the credentials used by IP2.  We therefore conclude that the two IPs must have been the same actor who downloaded and used the leaked SSH credentials. 

The IP addresses both resolve to the ``vpn.netspi.com'' domain, which belongs to NetSPI, a penetration testing, threat and attack surface management company~\cite{netspi}. 
While NetSPI used the SSH credentials found in TIAA's bucket, we do not find public evidence that TIAA is a customer of NetSPI.\footnote{While TIAA is not a public customer of NetSPI, both companies do have another connection: NetSPI's current CTO previously worked for TIAA as the Head of Cybersecurity Technology~\cite{venturebeat}.}
However, even if TIAA were a customer of NetSPI who authorized them to act on their behalf, neither the credentials we provided, nor the domain accessed belonged to TIAA.  Indeed, the downloaded informative document clearly stated that the SSH credentials belonged to a third-party contractor, and not to TIAA.  Notably, NetSPI did not append a three-digit identification number to the SSH password, which the informative document instructed to do. While it is conceivable that NetSPI understood that no access would be accomplished, their attempt at unauthorized access violates ethical norms.

\vspace{3pt}
\noindent
\textbf{SSH Brute Force.}\quad
Three unique actors---whose IP addresses belong to the high-fraud risk Packethub (AS\,147049), Midco VPN (AS,11232), and 31173 Services VPN (AS\,39351)---attempt all one thousand variations of the password leaked in the informative document. 

In the first brute force event, an actor used multiple IP addresses to find a company bucket, download the informative document, and attempt to log into the SSH Honeypot, across a 24-hour period.
We determined that two IP addresses used in these interactions belonged to the same actor. On October 30th, address IP1 listed the directory of the bucket ``oracledownload.''
Nearly 24 hours later, address IP2 attempted to download the informative document, DocA, in ``oracledownload'' that was originally presented to IP1.
However, IP2 was unsuccessful due to DocA having already been deleted because of our hourly updates (Section~\ref{sub:sub:sec:p2_buc_contents}).
Since IP1 was the only IP address to list the directory during DocA's existence, IP1 and IP2 must have been the same actor. 

We determined that this actor also used two additional IP addresses. Since the actor realized that the original DocA was no longer in the bucket, within 10 seconds a new address, IP3, listed the bucket directory again.
Yet another address, IP4, then downloaded the new informative document (DocB) originally listed to IP3. 
Having finally successfully downloaded DocB, which contained the leaked SSH credentials, IP1 attempted to login to the SSH Honeypot using the username and password displayed to IP4 five~minutes prior.  We therefore conclude that these four IP addresses were all the same actor that ultimately found and used the leaked SSH credentials. 
All four IPs belong to AS39351, ``31173 Services AB,'' which provides a VPN service~\cite{sweVPN}.

However, unlike the other SSH-abuse attempts, this actor clearly understood the requirement to append a three-digit identification number to the SSH password, as they iterated through \emph{all 1000 possibilities} while attempting to login. It is difficult to construct a credible scenario in which the actor could have believed these actions were authorized.\footnote{Moreover, since in most countries such unauthorized accesses are criminal acts (e.g., in the US under the Computer Fraud and Abuse Act, 18 USC 1030) this would be a substantial risk for a benign organization to take.} 

\subsection{Summary}
Actor engagement with company buckets increased with the presence of a vulnerability disclosure program (Section~\ref{sub:sub:sec:understand_comp}).
To identify scanners using VPNs and to better approximate unique actor activity, we developed a scalable approach to identify colluding IP addresses (Section~\ref{sub:sub:sec:actor_meth}).
Colluding IP addresses were more likely to participate in security critical behavior.
In the most abusive case, we identified eight separate events in which colluding IP addresses 
downloaded, read and understood our informative document, leading them to perform unauthorized login attempts into our honeypot server (Section~\ref{sub:sub:sec:ssh_events}).

%% file: tables/type_analysis_p2.tex
\begin{table}[t]
\caption{\textbf{Company Attribute Impact On Scanning}---%
\textnormal{Buckets that were named after companies with vulnerability disclosure programs (VDPs), or were in the technology sector, attracted statistically significantly more IP addresses.  Statistically significant increases of a metric, relative to all metrics with a smaller average value, are marked with a \texttt{*}.}
}
\centering
\begin{tabular}{lS[table-format=2.2]}
\toprule
  Company Attribute               &  \multicolumn{1}{r}{Avg IPs per bucket} \\
\midrule
 Has VDP                                 &         17.75*    \\
 No VDP                                 &               10.78  \\
\midrule
Technology                           &                28.19* \\
 Healthcare                             &     16.20     \\
 Transportation                 &           16.00       \\
 Financials                       &             15.69   \\
  Retail                                 &      10.25   \\
  Chem./Energy/Industrial  &          8.78 \\
  Eng./Construction/Materials   &            5.83 \\
 Aerospace/Defense                 &          5.50     \\
  Business Services                     &         4.00      \\

 



\bottomrule
\end{tabular}
\vspace{8pt}
\label{tab:typep2}
\end{table}

%% file: tables/top_10.tex
\begin{table}[t]
\caption{\textbf{Top 10 Most Scanned Buckets}---%
\textnormal{Over 58\% of the total number of IPs scanned the top 10 buckets, which spanned six companies. Every company, except Polaris, has a vulnerability disclosure program.}}
\centering
\begin{tabular}{ll}
\toprule
Bucket Name   & \# Unique IPs \\
\midrule
americanexpressdownload & 139 \\
americanexpressproduction  & 140 \\
oracledownload & 112\\
intuitproduction  & 55\\
oracleproduction  & 54\\
intuitdownload  & 46\\
nvidiadownload  & 44\\
nvidiaproduction & 44\\
targetdownload  & 43\\
polarisproduction  & 42\\
\midrule
Total &  719 (58.55 \%)\\
\bottomrule
\end{tabular}
\vspace{8pt}
\label{tab:top10}
\end{table}

%% file: ethics.tex
\section{Ethics}
\label{ethics}
We considered two classes of ethical issues in this work: potential human subject issues and potential harms to companies. As per discussions with our university's human subjects office, our work does not constitute human subjects research for the purposes of the US HHS Common Rule (45 CFR 46) because we are not collecting information about individuals.  Indeed, we have not solicited for any contact and any data that we receive is a byproduct of explicit actions taken to search for our buckets. We take no actions with this data (i.e., we do not act on the unauthorized logins to our infrastructure other than to log it) and thus we reason that any harms are minimal. This is consistent with a long line of research into third-party scanning behavior that has long been considered within the ethical norms of the community~\cite{PauleyNDSS2023}.

The second issue we considered was potential harm to companies due to trademark confusion or unwarranted reputational damage.  Working with the guidance of our university's general counsel, we designed our methodology to minimize these issues in several ways.  First, during the targeting phase of the study, we did not leak (i.e., advertise) buckets that included organizational names.  

Thus, organizational-named buckets could only be found by those actively and blindly guessing the names (minimizing any potential for confusion).  
Second, we provided multiple paths for resolving any confusion: including a contact email address in the informative document, a university affiliation that became clear if a visitor requested the access control list, and the normal notification path via AWS.  Finally, in the two instances in which we were contacted by brandholders (themselves notified by third parties, Section~\ref{sub:sub:sec:aws_reports}) we immediately removed the associated buckets as per their request.  

As well, we adhered to AWS terms of service, did not host any (real) sensitive data, and did not allow unauthorized login attempts from the attackers who sought to exploit our seemingly misconfigured buckets. 

%% file: discussion.tex
\section{Recommendations}
Our work demonstrates that buckets named after commercial entities are actively targeted and exploited.
Defending against cloud storage attackers is simple: configure buckets with sensitive information to be private. 
Unfortunately, prior work~\cite{continella2018there} has shown that thousands of buckets remain publicly exposed with sensitive information. 
Furthermore, over time, buckets are more likely to be misconfigured~\cite{cable2021stratosphere}. 
It is possible that misconfigured buckets are a symptom of owners who are unaware of the perils of exposing sensitive information, or are unaware that the bucket is theirs in the first place. 

Nevertheless, we propose the following recommendations for decreasing the risk of bucket exploitation, beyond simply making buckets private: 

\noindent 
\textbf{Scan assets.}\quad
Buckets named after the organization itself---as opposed to just low-entropy names---are the most attractive targets for scanners (Section~\ref{sub:sub:sec:buckets_found}).
We recommend that organizations consistently scan for both known and unknown cloud storage assets to immediately detect misconfigurations.
To scan for known assets, organizations can maintain a bucket bookkeeping system that periodically scans all buckets. 
To scan for unknown assets (e.g.,~\cite{assets_defense}), organizations should scan for ``easy-to-guess'' buckets named after the organization itself.
Scanning for unknown organization-named assets is not a new concept:
prominent organizations (e.g., Levi's, New Balance, etc) already use products (e.g. BrandShield~\cite{brandshield}) that protect brand reputation by scanning for organization-named Internet domains (e.g., `Levis.xyz') that might be engaging in phishing, fraud, or trademark infringement.
Thus, such brand-reputation scanning protections can, with likely minimal overhead, also scan for the presence of organization-named buckets to help protect organizations from data breaches (another threat against brand reputation).  
\looseness=-1

\noindent 
\textbf{Weigh risk according to organization type.}\quad
Companies with a VDP are more likely to be at risk than universities (Section~\ref{sub:sub:sec:understand_comp}). 
We recommend that security services which exist to help organizations trace stray and unknown assets (e.g., Censys~\cite{censys_assets}) weigh the risk of exposed bucket exploitation according to the organization type.
Such services can more aggressively scan and push to patch according to the organization type. 

\noindent 
\textbf{Use high entropy names.}\quad
High entropy names are less likely to be found and exploited because most scanners are simply guessing for bucket names (Section~\ref{sub:sub:sec:buckets_found}). 
Thus, organizations who create buckets with high entropy names (e.g., ``usaa-production-16219531'') are less likely to be found by the most common attacker strategy (albeit security through obscurity).

\noindent 
\textbf{Encourage cloud providers to protect customer assets.}\quad 
No matter the cause of misconfigured buckets, we believe cloud providers are in the best position to help trace and notify misconfigured bucket owners for two reasons.
First, while it is challenging for a third party to identify who truly owns a misconfigured bucket, cloud providers can  use the email address registered with the bucket owner's account to remind owners of buckets that are open to public access.
Second, while no public repository enumerating all existing buckets exists, cloud providers likely have some internal bookkeeping of resources that can be used to exhaustively identify all public buckets.
Currently, our experience with AWS is that it provided only limited proactive notifications:
across the 232~honeybuckets deployed across a total of 8~months, AWS only notified our account about four buckets that were publicly exposing sensitive data.

%% file: conclusion.tex
\section{Conclusion}
Actors scan and abuse the information found in cloud storage buckets. 
We deployed honeybuckets across two different experiments to measure how actors scan for buckets. 
Buckets named after companies---especially with a vulnerability disclosure program---were the most likely to be scanned and abused. 
Attackers constantly abused the permissions of the bucket they found: downloading files, uploading malicious executables, and even deleting existing content. 
Most concerning, we found that actors read and exploited the contents they downloaded; in eight cases, SSH login instructions leaked from our honeybuckets were precisely followed and used to attempt to gain unauthorized server access.
Given that attackers exploiting cloud storage is a reality, we hope our findings encourage both cloud storage operators and customers to track and secure their misconfigured buckets. 

%% file: data_avail.tex
\section*{Data Availability}
To promote reproducibility, upon request, we will share with researchers all raw data collected during this work's experiments.
Raw data will include logs of all scans that target any honeybucket or SSH-server.
We will also share the source-code used to identify colluding IPs. 

%% file: appendix.tex
\appendices
\section{Appendix}
\label{app}

We provide additional details regarding our honeybucket deployment, analysis, and results from Section~\ref{sec:phase1} and Section~\ref{sec:phase2}.
\input{tables/leakedMethod}
\input{tables/allOps}
\input{tables/uploadingObjects.tex}

\subsection{Bucket Scanning Ecosystem Experiment}
\input{tables/all_b_names}

\label{app:exp1}
In Section~\ref{sec:phase1}, we conducted an initial experiment to broadly study the bucket scanning ecosystem.
In Table~\ref{tab:leakedMeth}, we describe the platforms on which we leak our random alpha-numeric buckets. 
In Table~\ref{tab:p1bnames}, we list the exact names of all honeybuckets.
For brevity, we exclude the exact names of the 40 random-alphanumeric buckets that we leaked on various platforms.

In Table~\ref{tab:allOps}, we provide a general breakdown of the most popular operations performed by actors. Notably, 60\% of all operations only checked for the bucket's existence and 30\% only listed the files in a bucket without pursuing a further action (e.g., downloading a file).
In Table~\ref{table:uploaded_files}, we provide information about the 14~unique files that were uploaded by other users. We give the exact name of the file; an abbreviated description of its contents; whether the upload is a warning, malicious, or appears to serve no purpose; and the number of buckets the file appeared in. To protect researchers from any harm these unsolicited uploads may cause, researchers should download files in a protected and isolating computer environment to minimize the potential for harm introduced by these unsolicited uploads. However, simply listing the bucket contents can be done on a research machine.

\newpage
\subsection{Informative Document}
In Section~\ref{sec:phase2}, we conducted a follow-up experiment with additional lures, to understand to what extent will actors engage with the content they download.
We provide the exact text of the Informative Document, which is uploaded to all honeybuckets, below.\\
\label{appSSH}
\enlargethispage{-\baselineskip}
\begin{quote}
\begin{footnotesize}
\begin{verbatim}
%%%%%%%%%%%%%%%%%%%%%%%%%%%%%%%%%%%%%%%%%%%%%%%
Financial Analytics Storage Quick Start
%%%%%%%%%%%%%%%%%%%%%%%%%%%%%%%%%%%%%%%%%%%%%%%
This private storage bucket is used as a 
temporary  mechanism to transfer data from the 
on-premise  storage server to the client. 
Transaction data provided is only from the past 
three months. For  security reasons, each csv 
file is zip  encrypted. See Encryption Key 
Access tounzip. To provide the client with more 
historical data, an analysis consultant  must 
manually copy the data from the on-premise 
storage server to the storage bucket.
%%%%%%%%%
Login instructions:
%%%%%%%%%
To access the storage server, add the 
following entry to your ~/.ssh/config file.

Host transaction\_storage
Hostname 35.162.99.242
User bain\_fin\_analytics

Password Instructions:
Concatenate your unique 3-digit token 
with the secure numeric key, like so:
<token>62514653

If you do not have your unique 3-digit 
token, please ask your manager or email 
fanalytics.bain.IT@gmail.com.
%%%%%%%%%%%%
Encryption Key Access:
%%%%%%%%%%%%
Once logged into the server, navigate to
`/etc/s3\_zip\_encrypt`, in which the 
encryption key can be found.
%%%%%%%%%%%%%%%
Data transfer instructions:
%%%%%%%%%%%%%%%
Once logged into the server, navigate to 
`/mnt/transactions/<client name>/raw/`, 
in which the raw historical data will be found. 
Transferring data to and from the storage server
can be done through the following command:

`aws s3 sync /path/to/dir/ s3://COMPANY\_BUCKET'
%%%%%%%%%%%%%%%
Questions:
%%%%%%%%%%%%%%%
For all other questions, email 
fanalytics.bain.IT@gmail.com.
\end{verbatim}
\end{footnotesize}
\end{quote}
\newpage
\subsection{Fortune 500 Companies}
\input{tables/f500companies}
In Table~\ref{tab:f500}, we list the set of companies we name our buckets after from Section~\ref{sec:phase2}.
We separate buckets that do and do not have vulnerability disclosure programs, as well as indicate whether a monetary bug bounty is offered. 

\subsection{Identifying Colluding IPs}
\input{algorithms/algo_reformated.tex}
In Algorithm~\ref{fig:alg_collude}, we present the pseudocode to detect all cases of likely-colluding events from Section~\ref{sub:sub:sec:actor_meth}.
Algorithm~\ref{fig:alg_collude} takes in as input (1) an IP address
and (2) a token associated with the IP address. 
The algorithm outputs a set of likely colluding IPs.  

%% file: tables/leakedMethod.tex
\begin{table}[H]
\caption{\textbf{Leaking Bucket Names}---\textnormal{We leaked alphanumeric names of length~16 on a variety of platforms---public media, DNS, email---to measure if scanning actors are harvesting candidate names from passive sources.}}
\small
\begin{tabular}{ll}
\toprule
 Leaking service                   & How buckets  are leaked                          \\
\midrule
 Github repository                 & Both links in  README.md \\
 Pastebin                          & One link per paste     \\
 Twitter                          & One link    in a tweet            \\
Lab website               &  Both links in HTML               \\
 Google's DNS             & Query link's A Record        \\
 Spectrum's DNS           & Query link's A Record                          \\
 AT\&T's DNS             & Query link's  A Record       \\
 Open DNS  (Russian AS12714) &  Query link's  A Record    \\
 Open DNS  (Chinese AS4134) &      Query link's   A Record   \\
 Gmail drafts                      &   Both links in email   \\
\bottomrule
\end{tabular}
\label{tab:leakedMeth}
\end{table}

%% file: tables/allOps.tex
\begin{table}[H]
\caption{\textbf{Most Common Bucket Interactions}---\textnormal{The majority of bucket-IP pairs only checked for the bucket's existence, but did not pursue further action. Note, the respective bucket-IP pairs listed completed \textit{only} those operations.}}
\centering
\footnotesize
\begin{tabular}{lll}
\toprule
 Operation                                                              & Occurence           & Rolling         \\
 & (n=12233) & sum \\
\midrule
 Check bucket existence                                                   & 59.23\% (7246) & 59.23\%             \\
                                        
\midrule
 List bucket directory                                                    & 29.03\% (3551) & 88.26\%  \\
 \midrule
 Get object metadata                                                   & 2.49\% (305)  & 90.75\% \\
 \midrule
 1.Check bucket existence,                             & 1.51\%  (185)  & 92.26\% \\ 
 2. List bucket directory  & & \\
 \midrule
 Upload object                                                    & 1.37\%  (167)   & 93.63\%             \\
 \midrule
 Fail to download an object                                             & 0.89\%  (109)  & 94.52\%             \\
\midrule
 
 1.Check bucket existence,                                   & 0.83\%  (102)   & 95.35\%             \\
 2.Get ACL  && \\
\midrule
 1.List bucket directory,                           & 0.76\% (93)  & 96.11\%             \\
 2.Fail to download an object &   & \\
 \midrule
 Successfully download an object                                            & 0.62\%  (76)  & 96.73\%             \\
 \midrule
 1.Get ACL,                                    & 0.36\% (44)    & 97.09\%             \\
 2.Check bucket existence & & \\
\bottomrule
\end{tabular}
\label{tab:allOps}
\end{table}

%% file: tables/uploadingObjects.tex
\begin{table}[t]
\caption{\textbf{Uploaded Files With Content}---%
\textnormal{A total of 18 unique files were uploaded, of which only two were warning us that our buckets were misconfigured. 
Four files had malicious code and six files were inaccessible due to the uploader not permitting the read operation.  
Note that an additional 206 files were also uploaded, of which 91.3\% were of size zero.}}
\small
\centering
\begin{tabular}{llll}
\toprule
File Name & Description  & Category & \#  \\
& & & Up. \\
\midrule
`upload.png' & Warning  & Warn & 1 \\ 
`s3sec.txt' & Warning  & Warn & 1 \\ 
\midrule
`poc.jsp/' & reverse shell &  Mal & 5\\
`\_snapshot/test' & request  & Mal & 5 \\
& /etc/password &&\\
`\_snapshot/test2' & request & Mal & 5 \\
& /etc/password &&\\
`test-file.svg' &  re-direct to & Mal & 1 \\
& (ngrok.io)&&\\
\midrule
`bucket.png' & No access & - & 1 \\ 
`test' & No access  & - & 1 \\ 
`test.txt' & No access  & - & 1 \\
`indexx.html' & No access  & - & 1 \\
`hello.txt' & No access  & - & 1 \\
`s3-test.txt' & No access  & - & 1 \\
\midrule
`xss.svg' & image  & Benign & 1 \\
`xss1.svg' & image  & Benign & 1 \\
`Read.txt' & pen-test doc~\cite{s3secgithub}  & Benign & 1 \\
`testfile-nullg0re.txt' & ``this is a test''  & Benign & 4 \\
`test-test-nullg0re.txt' & ``this is a test'' & Benign & 1 \\
`testfile' & ``this is a test''  & Benign & 1 \\
\bottomrule
\end{tabular}
\label{table:uploaded_files}
\end{table}

%% file: tables/all_b_names.tex
\begin{table}[t]
\caption{\textbf{Bucket Names}---%
\textnormal{A list of bucket names from our first experiment from Section~\ref{sec:phase1}, in which we broadly studied the bucket scanning ecosystem. For brevity, we exclude the 40~alphanumeric leaked buckets.}}
\footnotesize
\begin{tabular}{ll}
\toprule
 Type        & Bucket name       \\
 \midrule
 Cryptocurrency                 &                  bitcoin-confidential  \\
                             &               bitcoin-secret             \\
                            &               ethereum-wallet              \\
                            &               ethereum-passwords              \\
\midrule
 Sensitive Keywords                          &                  passport10    \\
                             &               bank10             \\
\midrule
 Non-sensitive keywords                           &                  pretty10 \\
                             &              pictures10              \\
\midrule
DNSpop TGA                  &                  lyncdiscover     \\
                            &               612             \\
                            &               origin-www             \\
                            &               liboyulecheng            \\
\midrule
 Slurp TGA                &                  advogado  \\
                             &               applogie           \\
                            &               blognovo              \\
                            &               click1mail             \\
\midrule
 Pastebin TGA                        &                  screenshots-www       \\
                             &               www-slack              \\
                            &               www-download              \\
                            &               www-security              \\
\midrule
 Comparison Set                          &    confidentialfiles               \\
  (not in any TGA) &                 dont-open   \\
                            &                ignore-me           \\
                            &               pretty-pictures             \\
\midrule
Organization & teslaproduction, tesladownload\\
& teslapublic, teslaprivate \\
& teslasecurity, teslahidden \\
& walmartproduction, walmartdownload \\
& walmartpublic, walmartprivate \\
& walmartsecurity, walmarthidden \\
& tinderproduction, tinderdownload \\
& tinderpublic, tinderprivate \\
& tindersecurity, tinderhidden \\
& ucsdproduction \\
& ucsddownload \\
& ucsdpublic \\
& ucsdprivate \\
& ucsdsecurity \\
& ucsdhidden \\
& stanfordproduction \\
& stanforddownload \\
& stanfordpublic \\
& stanfordprivate \\
& stanfordsecurity \\
& stanfordhidden \\
& fbiproduction, fbidownload \\
& fbipublic, fbiprivate \\
& fbisecurity, fbihidden \\
& ciaproduction, ciadownload \\
& ciapublic, ciaprivate  \\
& ciasecurity, ciahidden\\
& nypdproduction, nypddownload  \\
& nypdpublic, nypdprivate \\
& nypdsecurity, nypdhidden \\
\bottomrule
\end{tabular}
\label{tab:p1bnames}
\end{table}

%% file: tables/f500companies.tex
\begin{table*}[t]
\caption{\textbf{Company Names}---%
\textnormal{The final list of companies we used in naming our buckets, with each name concatenated with `download' and `production' (e.g. 3mproduction, 3mdownload). The names indicated with a * were originally Comcast and Equinix. In the vulnerability disclosure program column we cite the source explaining the company's vulnerability disclosure program and indicate with a \$ if the company has a monetary bug bounty.
}}
\small
\centering
\begin{tabular}{c|c|c|c|c}
\toprule
\multicolumn{3}{c}{Vulnerability Disclosure Program} &  \multicolumn{2}{c}{No Vulnerability Disclosure Program}                           \\
\cmidrule(r){1-3}\cmidrule(r){4-5}
Name & Rank& Vulnerability  & Name & Rank \\ 
& & Disclosure  & &  \\
& & Program & &  \\
\midrule
CVS Health & 4 & \$ \cite{cvs} & Raytheon Technologies & 58\\
United Health Group & 5 & \cite{united}& Charter Communications & 69\\
Target & 32&\cite{target} & Tyson Foods &  81\\
State Farm Insurance  & 42&\cite{state} & 3M & 102\\
Pfizer & 43& \cite{pfizer} &  Applied Materials & 156\\
General Electric & 48&\cite{general}& Lithia Motors & 158 \\ 
 Goldman Sachs Group & 57& \$ \cite{goldman} & Hartford Financial Services & 160\\
HCA Healthcare & 62 & \cite{hca}& Lincoln National & 187\\
Deere & 84 & \cite{deere}& Wesco International & 200\\
American Express & 85 & \cite{amex_vdp} & L3 Harris Technologies & 206\\
TIAA & 90& \cite{tiaa} & Automatic Data Processing & 242\\
*Oracle & 91& \cite{oracle} & Pioneer Natural Resources&  248\\ 
 USAA &96 &\$ \cite{usaa} & Pulte Group & 267 \\
Northwestern Mutual & 97 & \cite{northwest}  & Oreilly Automotive&  279\\ 
Capital One Financial & 108 & \cite{capital} & Rocket Companies & 282 \\
Nvidia & 134 & \cite{nvidia} & Vistra & 315\\
PNC Financial Services & 178& \cite{pnc} & Unum Group & 317 \\
Charles Schwab & 188& \cite{charles} & Altice USA & 355 \\
Otis Worldwide & 254 & \cite{otis} & ODP & 379 \\
Discover Financial Services & 281& \cite{discover} & Delek US Holdings & 346\\
Carvana & 290& \cite{carvana}  & Univar Solutions & 369 \\
Tractor Supply & 294& \cite{tractor} & Burlington Stores & 377 \\
Keurig Dr Pepper & 296& \cite{keurig} & Jefferies Financial Group & 387\\
CSX & 298& \cite{csx} & Polaris & 419 \\ 
Boston Scientific & 319& \cite{boston} & MasTec & 429 \\
 Booking Holdings &340 & \$ \cite{booking} & GXO Logistics & 430\\
*Intuit & 366& \cite{intuit} & Westinghouse Air Brake Tech & 439\\
Dover & 433& \cite{dover} & Hertz Global Holdings &  462\\
Analog Devices & 463& \cite{analog}  & Graphic Packaging Holding & 466 \\
Regions Financial & 489& \cite{regions} & Landstar System & 491 \\
\bottomrule
\end{tabular}
\label{tab:f500}
\end{table*}

%% file: algorithms/algo_reformated.tex
\begin{algorithm}[t]
\begin{footnotesize}
\SetAlgoLined
\tcp{extract filename from token}
\If{ token==valid SSH password} { 
file = unhash(token)
}
\uElseIf{ token==informative doc} { 
file = token
}
\Else{return \tcp{IP did not download a file or SSH}} 
\tcp{base case: no collusion}
\If{$\mbox{time(`IP lists directory')} < \mbox{time(`IP downloads file')}$}
{return IP \tcp{the IP listed directory for itself}}
\tcp{get time of first download}
first\_download=MIN(logs[logs[`ips']==IP][`time']) \\
\tcp{get bucket of first download}
bucket=logs[logs[`time']==first\_download][`bucket\_name']\\
\tcp{find if download was successful}
successful\_download=\\
logs[logs[`time']==first\_download][`error']\\
\tcp{get upload time}
upload\_time=get\_upload\_time(file)\\
\tcp{get deletion time}
delete\_time=upload\_time + 1 hour\\
\If{successful\_download}
{cut\_off=first\_download \tcp{file not deleted by download time}}
\If{!successful\_download}
{cut\_off=delete\_time \tcp{file already deleted by download time}}
\tcp{get all directory-listing IPs }
ips\_list\_dir= logs[logs[`operation']==`list directory'][`ips']\\
\For{i in ips\_list\_dir}
{\tcp{if directory listing between upload and cut off time, mark colluding\_ips}
\If{logs[logs[`ips']==i][`time'] $\geq$ upload\_time \& logs[logs[`ips']==i][`time'] $\leq$ cut\_off}
{colluding\_ips.append(i)}}

return colluding\_ips \\
  \caption{find\_colluding\_ips(IP,token):}
  \label{fig:alg_collude}
 \end{footnotesize}
\end{algorithm}